\documentclass[12pt]{iopart}
\usepackage{amssymb}
\usepackage{graphicx}
\usepackage{dcolumn}
\usepackage{enumerate}
\usepackage{bm,bbm}

\expandafter\let\csname equation*\endcsname\relax
\expandafter\let\csname endequation*\endcsname\relax

\usepackage{amsmath}

\newcommand{\av}[1]{\langle #1\rangle}

\begin{document}

\title[Heat transport in the $XXZ$ spin chain]{Heat transport in the $XXZ$ spin chain: from ballistic to diffusive regimes and dephasing enhancement}

\author{J J Mendoza-Arenas$^1$,  S Al-Assam$^1$, S R Clark$^{2,1}$\\ and D Jaksch$^{1,2}$}

\address{$^1$Clarendon Laboratory, University of Oxford, Parks Road, Oxford OX1 3PU, United Kingdom}
\address{$^2$Centre for Quantum Technologies, National University of Singapore, 3 Science Drive 2, Singapore 117543}

\ead{j.mendoza-arenas1@physics.ox.ac.uk}

\begin{abstract}
In this work we study the heat transport in an $XXZ$ spin-$1/2$ Heisenberg chain with homogeneous magnetic field, incoherently driven out of equilibrium by reservoirs at the boundaries. We focus on the effect of bulk dephasing (energy-dissipative) processes in different parameter regimes of the system. The non-equilibrium steady state of the chain is obtained by simulating its evolution under the corresponding Lindblad master equation, using the time evolving block decimation method. In the absence of dephasing, the heat transport is ballistic for weak interactions, while being diffusive in the strongly-interacting regime, as evidenced by the heat-current scaling with the system size. When bulk dephasing takes place in the system, diffusive transport is induced in the weakly-interacting regime, with the heat current monotonically decreasing with the dephasing rate. In contrast, in the strongly-interacting regime, the heat current can be significantly enhanced by dephasing for systems of small size.
\end{abstract}

\section{Introduction} \label{intro}
The transport properties of low-dimensional interacting quantum systems have been the object of intense research for several years. The development of powerful numerical methods have allowed important insights into the conduction regimes of these systems for different interaction strengths to be obtained \cite{gobert2005real,langer2009real,prosen2009matrix,kajala2011prl,langer2011prb}. In addition, the ground-breaking advances in the manipulation of cold atom systems have lead to the recent experimental observation of interesting transport phenomena, which strongly depend on the interaction between particles \cite{schneider2012nat,errico2012,brantut2012sci,stadler2012nat,ronzheimer2013arxiv}. In spite of this vast theoretical and experimental progress, a complete picture of the transport properties of interacting quantum systems is still lacking.\\The analysis of transport phenomena becomes more involved when its coupling with the environment is considered. The study of the competition between this unavoidable incoherent coupling and coherent dynamics, and its consequences on transport effects, has received much attention recently due to the observation of long-lived quantum coherences in light-harvesting complexes even at room temperature \cite{engel2007nat,lee2007sci,collini2009sci,schlau2012nat,scholes2011nat}. It has been observed that dephasing processes can enhance the transfer of particles \cite{plenio2008njp,mohseni_jcp2008,deph_assisted2,deph_assisted_plenio} and heat \cite{manzano2013plos} in some non-interacting 2D networks, as well as in 1D chains for non end-to-end conduction \cite{kassal2012njp}; the optimal transport then occurs at an appropriate balance between coherent and incoherent effects. Recently we have shown that in a 1D chain with end-to-end transport, a large enhancement of spin current emerges due to bulk dephasing, if strong interactions exist \cite{we}\footnote{The model considered in Ref. \cite{we} corresponds to a chain of interacting spinless fermions, which under a Jordan-Wigner transformation can be mapped to a spin chain with $XXZ$ interactions.}. Since dephasing processes take place in each site of the system, and their corresponding jump operators do not commute with the Hamiltonian, they not only randomize phase information between spins, but also dissipate energy. For interactions strong enough to produce separated bands of eigenstates of low mobility \cite{Sutherland,winkler2006nat,haque2010self}, dissipation induces transitions from such bands of large potential energy to scattering states with larger kinetic energy, thereby enhancing the spin current. This general mechanism of environment-assisted transport, described in Ref. \cite{we}, explains some recent experimental results, in which the presence of noise significantly enhances the expansion of a strongly-interacting bosonic gas in a disordered potential \cite{errico2012}. Other interesting dynamical consequences of the interplay between interactions and dissipation, e.g. interaction impeded decoherence \cite{kollath2012prl,kollath2012arxiv} and glass-like disorder-free states \cite{poletti2012arxiv} have been lately reported.\\Experimentally, a commonly-used method to study transport properties of atomic quantum gases consists of releasing the atomic ensemble from its trap and observing its expansion under different conditions such as particle-particle interactions, disorder, noise, etc. \cite{schneider2012nat,errico2012,ronzheimer2013arxiv}; several numerical simulations have been performed for similar situations, in both cold atomic gases \cite{kajala2011prl} and spin systems \cite{gobert2005real,langer2009real,langer2011prb}. A different method consists of analyzing the currents of systems driven out of equilibrium by asymmetric reservoirs at the boundaries \cite{prosen2009matrix,benenti2009charge,michel2008prb,wu2011prb,popkov2012pre,bruderer2012pra,kollath2012thermo,manzano2012pre}. A very important step towards the experimental implementation of such configurations in atomic gases has been achieved recently, where the transport of ultracold fermions through a mesoscopic channel connecting two particle reservoirs has been realized \cite{brantut2012sci,stadler2012nat}. The incorporation of strong interactions by Feshbach resonances and of external noise by an appropriate modulation of an underlying optical lattice \cite{errico2012} is feasible given the recent advances in the control of ultracold atoms.\\An archetypical model to analyze the interplay of interactions and environmental effects on transport processes is given by the 1D $XXZ$ Hamiltonian, which is one of the simplest and most studied models of interacting quantum systems. This model can be experimentally realized in systems such as optical cavities \cite{kay2008reproducing}, ultracold atoms in optical lattices \cite{simon2011nat} and some cuprate materials, with ${\mathrm{SrCuO}}_2$ and $\mathrm{Sr}_2\mathrm{CuO}_3$ the most prominent  realizations of the isotropic model \cite{motoyama1996prl,ami1995prb,suzuura1996prl,zaliznyak2004prl}, and with (NO)Cu(NO$_3$)$_3$ \cite{janson2010prb} and LiCuVO$_4$ \cite{krug2002prb} proposed to feature strong interactions. The commonly accepted picture of spin transport in this model, obtained from recent theoretical studies, corresponds to ballistic and diffusive behavior in the weakly- and strongly-interacting regimes, respectively \cite{gobert2005real,langer2009real,prosen2009matrix,benenti2009charge,znidaric_pertur_delta}. Theoretical analysis, based on the integrability of the $XXZ$ model \cite{Sutherland} and the conservation of the heat current \cite{zotos1997prb}, has led to the conclusion that for any interaction strength between spin excitations the thermal Drude weight of the system is nonzero. Consequently the heat conductivity is infinite, i.e., the heat transport is ballistic \cite{zotos1997prb}. This result agrees with several analytical \cite{louis2003prb,shimshoni2003prb,rozhkov2005prl,chernyshev2005prb,heidrich2005prb,louis2006prb,heidrich2007epjst,lifa2008prb,shimshoni2009prb} and numerical \cite{langer2011prb,ajisaka2012prb} studies, and has been supported experimentally by the measurement of large heat conductivities in cuprates \cite{hlubek2010prb,hlubek2012jsm,kawamata2010jpcs}.\\Interestingly, the recent theoretical studies on non-equilibrium $XXZ$ chains using boundary reservoirs suggest that spin and heat transport properties differ from this generally accepted picture. A transition from ballistic to diffusive heat transport for strong interactions between spin excitations was observed in a chain with a strong magnetic field \cite{michel2008prb}. In a similar setup with lower fields, ballistic heat and spin transport were observed for the non-interacting regime only, arguing that finite interactions induce diffusive transport \cite{wu2011prb}. Therefore more effort is still required to fully understand the transport properties in these driven systems.\\To help establish a complete picture of transport phenomena in quantum systems where interactions and dephasing have a prominent role, in the present work we study the heat current through an $XXZ$ chain driven out of equilibrium by unequal reservoirs at the boundaries, and with dephasing processes in the bulk. Initially we present the model to be analyzed and the definition of the local heat current in Section \ref{model}. We then study the heat transport properties of the $XXZ$ spin chain for several parameter regimes. In Section \ref{nodephasing} we characterize the nature of heat transport through the system in a homogeneous magnetic field without dephasing. We find that heat propagates ballistically in the weakly-interacting regime, while for strong interactions, negative differential conductivity (NDC) emerges, with diffusive behavior at weak driving only and an insulating state at large driving. Then we observe the effects of bulk dephasing on heat transport in the different interaction regimes. In Section \ref{depheffects1} we consider the case of weak interactions. By analogy with Refs. \cite{znidaric2010dephasing,znidaric2011solvable} we show that dephasing induces a non-equilibrium phase transition from ballistic to diffusive regimes, with the heat transport monotonically degraded by dephasing. For strong interactions, studied in Section \ref{depheffects1}, dephasing degrades the NDC effect, leading to a diffusive behavior of the system for all driving strengths. In addition, the heat current can be enhanced by moderate dephasing rates for small systems.

\section{Model of heat transport in a 1D Heisenberg chain} \label{model}
\subsection{Non-equilibrium $XXZ$ spin chain} \label{model_subsection}
\begin{figure}
\begin{center}
\hspace{1cm}
\includegraphics{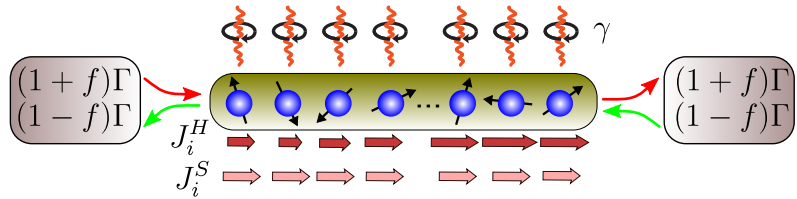}
\hspace{-1cm}
\caption{\label{fig1} Scheme of a spin chain driven out of equilibrium by the action of reservoirs at the edges, with coupling strength $\Gamma$ and driving parameter $f$. The dephasing processes have a rate $\gamma$. The light arrows indicate the homogeneous flow of spin excitations. The dark arrows represent the heat current, which varies through the chain due to the energy dissipation; the amplitudes of the latter are based on the results shown in Fig. \ref{fig6}(b).}
\end{center}
\end{figure}

We consider the nearest-neighbor interacting spin-$\frac{1}{2}$ $XXZ$ spin chain, in the presence of a homogeneous magnetic field. The Hamiltonian is given by (taking $\hbar=1$)
\begin{equation} \label{hami}
H = \tau\sum_{j=1}^{N-1}(\sigma_j^x\sigma_{j+1}^x+\sigma_j^y\sigma_{j+1}^y+\Delta\sigma_j^z\sigma_{j+1}^z)+B\sum_{j=1}^{N}\sigma_j^z,
\end{equation}
with $\sigma_i^\alpha$ ($\alpha=x,y,z$) the Pauli matrices at lattice site $i$, $N$ the number of sites, $\tau$ the exchange coupling between nearest neighbors (we set the energy scale by taking $\tau=1$ from here on), $B$ the magnetic field and $\Delta$ the anisotropy parameter. The $\sigma_j^z\sigma_{j+1}^z$ terms act as an interaction penalizing the alignment of neighboring spins, while the in-plane terms cause spin-flip ``hopping'' processes. The ground-state phase diagram of this model is well known: in the absence of magnetic field, $\Delta<-1$ and $\Delta>1$ correspond to ferro- and antiferromagnetic (gapped) phases, respectively, and $-1<\Delta<1$ to a gapless state. A finite $B$ shifts the critical points to a larger interaction strength $|\Delta|$ if $\Delta>0$, and to a weaker strength $|\Delta|$ if $\Delta<0$; see Ref. \cite{takahashi} for a detailed description. In the present work we will only consider interactions $\Delta>0$, and magnetic field $B>0$.\\To model a non-equilibrium configuration we introduce Markovian reservoirs. The dynamics of the system is captured by a Lindblad quantum master equation \cite{breuer}
\begin{equation} \label{master_eq}
 \frac{\partial\rho}{\partial t}=-i[H,\rho] + \mathcal{L}(\rho),
\end{equation} 
where $\rho$ is the density matrix of the chain and $\mathcal{L}(\rho)$ is the dissipator describing its coupling to the environment. In Lindblad form,
\begin{equation}
\mathcal{L}(\rho)=\sum_{k}\mathcal{L}_k(\rho)=\sum_{k}\biggl(L_{k}\rho L_{k}^{\dagger}-\frac{1}{2}\{L_{k}^{\dagger}L_{k},\rho\}\biggr),
\end{equation}
where $L_k$ are the jump operators corresponding to the coupling of the system to the environment, and $\{.,.\}$ is the anti-commutator. We consider three different couplings $\mathcal{L}(\rho)=\mathcal{L}_{\rm L}(\rho)+\mathcal{L}_{\rm d}(\rho)+\mathcal{L}_{\rm R}(\rho)$, described below.\\The dissipators $\mathcal{L}_{\rm L}(\rho)$ and $\mathcal{L}_{\rm R}(\rho)$ represent the boundary driving of the system to a non-equilibrium configuration, and correspond to the jump operators \cite{benenti2009charge}
\begin{equation} \label{driv_oper}
L_{\rm L,R}^+=\sqrt{\Gamma(1\pm f)/2}\,\sigma_{1,N}^+\quad L^-_{\rm L,R}=\sqrt{\Gamma(1\mp f)/2}\,\sigma_{1,N}^-,
\end{equation}
with $\sigma_j^{\pm}=\frac{1}{2}(\sigma_j^x\pm i\sigma_j^y)$. These operators create and annihilate spin excitations at the boundary sites of the system, with $f$ the driving strength ($0\leq f \leq1$) and $\Gamma$ the coupling strength to the spin reservoirs (we take $\Gamma=1$). The driving strength $f$ controls the non-equilibrium forcing, and in isolation would impose unequal average spin polarizations in the boundaries, $\av{\sigma^z_1} =  f$ and $\av{\sigma^z_N} =  -f$ \cite{znidaric2011solvable}. A zero bias $f=0$ leads to a steady state $\rho = \mathbbm{1}/2^N$ \cite{burgarth2007prl}, indicating that the system is formally at infinite temperature; no spin nor heat current exists in this case since spin-flip excitations are created and annihilated at the same rate on both boundaries. At finite driving $f$ a net spin current flow is established due to the asymmetry between the two edges. At maximum bias $f=1$ the boundary jump operators reduce to $\sigma^+_1$ and $\sigma^-_N$.\\The bulk dephasing processes are described by the dissipator $\mathcal{L}_{\rm d}(\rho)=\sum_{j=1}^N\mathcal{L}^{(j)}_{\rm d}(\rho)$, with a homogeneous dephasing rate $\gamma$ and jump operator at site $j$
\begin{equation} \label{deph_oper}
L_{j}^{\rm d} = \sqrt{\gamma}\sigma_{j}^{z}.
\end{equation}
A schematic representation of the system, including the different incoherent processes, is shown in Fig. \ref{fig1}.\\
The interplay between coherent and incoherent phenomena leads to a non-equilibrium steady state (NESS) $\rho_{\rm ss}$, which satisfies $\partial\rho_{\rm ss}/\partial t=0$. This state is thus identified by the stationary values of its spin and heat currents. Since the Hilbert space containing $\rho_{ss}$ grows exponentially with the size of the system (as $4^N$), it is not possible to obtain the exact steady state for large $N$ numerically. In addition, a simple analytical approximation to the steady state of a large system is only feasible in the non-interacting regime \cite{znidaric_pertur_delta,znidaric2011solvable}. To obtain $\rho_{\rm ss}$, we then directly simulate Eq. \eqref{master_eq} and take the long time limit applying the time evolving block decimation (TEBD) method \cite{zwolak2004mixed,cirac2004prl} to a matrix product operator description of $\rho(t)$. Our implementation of the algorithm is based on the open source Tensor Network Theory (TNT) library \cite{tnt}, and allows us to analyze efficiently the transport response of strongly-interacting systems larger than those considered in previous studies \cite{michel2008prb,wu2011prb}.

\subsection{Alternative driving schemes}
In this work we study the heat transport induced by a magnetization imbalance at the boundaries of a spin chain, imposed by the Lindblad driving processes of Eq.~\eqref{driv_oper}. This means that we actually observe the magnetothermal behavior of the system, instead of the direct response to an energy imbalance enforced at the boundaries (see also \cite{popkov2013njp}). To analyze the latter situation, different driving mechanisms have been proposed. One possibility consists on using two-spin driving operators \cite{prosen2009matrix}, designed to induce a Gibbs state of determined temperature when applied to an isolated pair of interacting spins (analogous to the single-spin operators of Eq.~\eqref{driv_oper} which induce a state with magnetization $f$ on an isolated spin). Another recently proposed configuration consists of connecting the chain to finite meso-reservoirs at the boundaries, driven to a state of fixed temperature and chemical potential by a Lindblad-type coupling to super-reservoirs \cite{ajisaka2012prb}.\\These alternative and somewhat complicated driving processes would allow a complete study of the heat current directly generated by an energy imbalance imposed at the boundaries of the system, as well as of its magnetothermal response, for different interacting regimes. Here we instead consider the much simpler and intuitive driving scheme of Eq.~\eqref{driv_oper}, which still allows the observation of interesting phenomena in various parameter regimes. The transport properties of strongly-correlated systems under the alternative driving schemes are the subject of current research \cite{we2}.

\subsection{Spin and heat currents} \label{defincurrents}
In our previous work \cite{we}, we analyzed the spin transport properties of the system described above, by computing the spin current and local magnetization. The spin current operator follows from the continuity equation, and is given by
\begin{equation}
J^{\rm S}_i=2(\sigma_i^x\sigma_{i+1}^y-\sigma_i^y\sigma_{i+1}^x),\quad i=1,\dots,N-1,
\end{equation}
where its expectation value is spatially homogeneous in the steady state, so $\langle J^{\rm S}_i\rangle_{\rm ss}=\Tr(\rho_{\rm ss}J^{\rm S}_i)=J^{\rm S}$ for all $i$\footnote{In the present work we are interested in the transport properties in the steady state of the system. Hereafter all the expectation values are calculated for such a state, and for simplicity we will drop the subindex ${\rm ss}$ indicating steady state expectation values. Nevertheless the definitions of heat currents presented in Section \ref{defincurrents} remain valid in the transient regime.}. The spin current $J^{\rm S}$ and the magnetization profile indicate different spin transport regimes. The ballistic regime, corresponding to $\Delta<1$ and $\gamma=0$, is characterized by a nearly-flat magnetization profile, and by a current proportional to the driving and independent of the size of the system, i.e. $J^{\rm S}\propto f$ \cite{znidaric2011solvable,znidaric2010dephasing}. In the diffusive regime, corresponding to $\Delta>1$, weak driving and $\gamma=0$, as well as to any $\Delta$ and $\gamma>0$ \cite{we,znidaric2010dephasing}, the system satisfies the diffusion equation
\begin{equation}
J^{\rm S}=\kappa_{\rm S}\nabla S,
\end{equation}
with $\kappa_{\rm S}$ the spin conductivity and $\nabla S=(\langle\sigma_1^z\rangle-\langle\sigma_N^z\rangle)/(N-1)$ the magnetization gradient. This transport regime is characterized by a linear magnetization profile and a spin current scaling $J^{\rm S}\propto f/N$ \cite{we,benenti2009charge}.\\In the present paper we focus on heat transport. Due to the energy dissipation caused by dephasing, the heat current is not homogeneous in the steady state, but different at every site. We denote by $J^{\rm H}_i$ the heat current operator at site $i$; to obtain its expectation value, we first write the Hamiltonian in the form
\begin{equation}
H=\sum_{i=1}^{N-1}\varepsilon_{i,i+1}=\sum_{i=1}^{N-1}(h_{i,i+1}+b_{i,i+1}),
\end{equation}
where $h_{i,i+1}$ and $b_{i,i+1}$ correspond to the $XXZ$ coupling and magnetic field terms, respectively, and are given by
\begin{equation}
h_{i,i+1}=\sigma_i^x\sigma_{i+1}^x+\sigma_i^y\sigma_{i+1}^y+\Delta\sigma_i^z\sigma_{i+1}^z,\quad b_{i,i+1}=\frac{B}{2}[\sigma_i^z(1+\delta_{i,1})+\sigma_{i+1}^z(1+\delta_{i+1,N})],
\end{equation}
where $\delta$ denotes the Kronecker delta.\\To obtain the heat current, we calculate the rate of change of the local energy density $\varepsilon_{i,i+1}$. From the master equation \eqref{master_eq}, we have
\begin{align}
\begin{split}
\fl\frac{\partial\langle\varepsilon_{i,i+1}\rangle}{\partial t}=\Tr\biggl(\varepsilon_{i,i+1}\frac{\partial\rho}{\partial t}\biggr)=&-i\Tr(\varepsilon_{i,i+1}[H,\rho])+\Tr(\varepsilon_{i,i+1}\mathcal{L}_{\rm L}(\rho))\\&+\Tr(\varepsilon_{i,i+1}\mathcal{L}_{\rm R}\rho))+\sum_{j=1}^N\Tr(\varepsilon_{i,i+1}\mathcal{L}_{\rm d}^{(j)}(\rho)),
\end{split}
\end{align}
where $i=1,\ldots,N-1$. In the steady state, the previous equation is reduced to
\begin{align} \label{heatcurr1}
\begin{split}
\fl\frac{\partial\langle\varepsilon_{i,i+1}\rangle}{\partial t}&=i\langle[H,\varepsilon_{i,i+1}]\rangle+\Tr(\varepsilon_{1,2}\mathcal{L}_{\rm L}(\rho))\delta_{i,1}+\Tr(\varepsilon_{N-1,N}\mathcal{L}_{\rm R}(\rho))\delta_{i+1,N}\\&+\Tr(\varepsilon_{i,i+1}\mathcal{L}_{\rm d}^{(i)}(\rho))+\Tr(\varepsilon_{i,i+1}\mathcal{L}_{\rm d}^{(i+1)}(\rho))=0.
\end{split}
\end{align}
Next we consider the continuity equation at each pair of neighboring sites in the steady state
\begin{equation}  \label{heat_continuity}
\frac{\partial\langle\varepsilon_{i,i+1}\rangle}{\partial t}=-\nabla\langle J_i^{\rm H}\rangle+\frac{\partial\langle\varepsilon_{i,i+1}\rangle}{\partial t}\biggl|_{\text{Env}}=0,
\end{equation}
where $\nabla\langle J_i^{\rm H}\rangle=\langle J_{i+1}^{\rm H}\rangle-\langle J_i^{\rm H}\rangle$ is the gradient of the heat current, and $(\partial\langle\varepsilon_{i,i+1}\rangle/\partial t)|_{\text{Env}}$ is the rate of change of the local energy density due to the coupling of the system with the environment. By comparing Eqs. \eqref{heatcurr1} and \eqref{heat_continuity}, we identify
\begin{equation} \label{totalcurrent}
i\langle[H,\varepsilon_{i,i+1}]\rangle=-\nabla\langle J_i^{\rm H}\rangle=\langle J_i^{\rm H}\rangle-\langle J_{i+1}^{\rm H}\rangle\quad\rightarrow\quad \langle J_i^{\rm H}\rangle=i\langle[\varepsilon_{i-1,i},\varepsilon_{i,i+1}]\rangle,\quad 2\leq i\leq N-1.
\end{equation}
Note that this definition of the heat current does not apply to the boundaries of the chain. For such cases, the heat current results from the action of the left and right reservoirs. If we take
\begin{equation} \label{continuity_boundaries}
\langle J_{1}^{\rm H}\rangle=\Tr(\varepsilon_{1,2}\mathcal{L}_{\rm L}(\rho))\qquad-\langle J_N^{\rm H}\rangle=\Tr(\varepsilon_{N-1,N}\mathcal{L}_{\rm R}(\rho)),
\end{equation}
then we have a consistent definition of the heat current over the entire system, since the continuity equation for any pair of nearest neighbors gives
\begin{equation}
\langle J_{i+1}^{\rm H}\rangle-\langle J_i^{\rm H}\rangle=\Tr(\varepsilon_{i,i+1}\mathcal{L}_{\rm d}^{(i)}(\rho))+\Tr(\varepsilon_{i,i+1}\mathcal{L}_{\rm d}^{(i+1)}(\rho))=-4\gamma\langle\sigma_{i}^x\sigma_{i+1}^x+\sigma_{i}^y\sigma_{i+1}^y\rangle.
\end{equation}
As expected, the difference between heat currents at neighboring sites arises solely from the energy dissipation due to dephasing processes at those sites.\\In order to understand the features of the heat transport we separate the heat current in Eq. \eqref{totalcurrent} into two contributions: that arising from $XXZ$ interactions only, given at site $i$ ($2\leq i\leq N-1$) by
\begin{equation} \label{xxz}
\begin{split}
\fl\langle J_i^{\rm XXZ}\rangle&=i\langle[h_{i-1,i},h_{i,i+1}]\rangle=2\langle(\sigma_{i-1}^y\sigma_i^z\sigma_{i+1}^x-\sigma_{i-1}^x\sigma_i^z\sigma_{i+1}^y)\\&+\Delta(\sigma_{i-1}^z\sigma_i^x\sigma_{i+1}^y-\sigma_{i-1}^y\sigma_i^x\sigma_{i+1}^z)+\Delta(\sigma_{i-1}^x\sigma_i^y\sigma_{i+1}^z-\sigma_{i-1}^z\sigma_i^y\sigma_{i+1}^x)\rangle,
\end{split}
\end{equation}
and the heat current carried by the net flow of spin excitations, corresponding to 
\begin{equation} \label{mag}
\begin{split}
\fl\langle J_i^{\rm B}\rangle&=i\langle[\varepsilon_{i-1,i},\varepsilon_{i,i+1}]\rangle-i\langle[h_{i-1,i},h_{i,i+1}]\rangle\\&=B\langle(\sigma_{i-1}^x\sigma_i^y-\sigma_{i-1}^y\sigma_i^x)+(\sigma_i^x\sigma_{i+1}^y-\sigma_i^y\sigma_{i+1}^x)\rangle=\frac{B}{2}\langle(J^{\rm S}_{i-1}+J^{\rm S}_i)\rangle
\end{split}
\end{equation}
We define similar contributions at the boundaries. For site $i=1$,
\begin{equation}
\langle J_1^{\rm XXZ}\rangle=\Tr(h_{1,2}\mathcal{L}_{\rm L}(\rho)),\qquad \langle J_1^{\rm B}\rangle=\Tr(b_{1,2}\mathcal{L}_{\rm L}(\rho)),
\end{equation}
where we obtain
\begin{equation}\label{boundary1}
\langle J_1^{\rm XXZ}\rangle = -\Gamma(\langle h_{1,2}\rangle+\Delta\langle\sigma_1^z\sigma_2^z\rangle)/2+\Gamma f\Delta\langle\sigma_{2}^z\rangle, \qquad \langle J_1^{\rm B}\rangle = \Gamma B(f-\langle\sigma_1^z\rangle).
\end{equation}
The corresponding definitions for site $N$ are
\begin{equation}\label{boundary2}
\langle J_N^{\rm XXZ}\rangle = \Gamma(\langle h_{N-1,N}\rangle+\Delta\langle\sigma_{N-1}^z\sigma_N^z\rangle)/2+\Gamma f\Delta\langle\sigma_{N-1}^z\rangle, \qquad \langle J_N^{\rm B}\rangle =\Gamma B(f+\langle\sigma_N^z\rangle).
\end{equation}
The total heat current operator at site $i$ is then $J_i^{\rm H}=J_i^{\rm XXZ}+J_i^{\rm B}$. Since the spin current of the steady state of the system is homogeneous, so is $J_i^{\rm B}$; we thus define $J^{\rm B}=\langle J_i^{\rm B}\rangle=BJ^{\rm S}$. The spatial variation of the heat current arising from dephasing processes is then entirely contained within $\langle J_i^{\rm XXZ}\rangle$.\\Note from Eqs. \eqref{boundary1} and \eqref{boundary2} that the spin current $J^{\rm S}$ and the homogeneous component of the heat current $J^{\rm B}$ can be directly obtained from the magnetization at the boundaries. This was previously found for the case $\Delta=0$ \cite{znidaric2011solvable}. Now we observe that the same applies for all interaction strengths $\Delta$, and that $J^{\rm S}$ and $J^{\rm B}$ are given by the deviation of the magnetization of the boundary spins from the one that would be imposed by the left and right reservoirs to isolated spins ($\pm f$).\\We now study the total heat current and its components in the non-equilibrium setup described in Section \ref{model_subsection}, and observe its behavior in the weakly ($\Delta<1$) and strongly ($\Delta>1$) interacting regimes, in the absence and presence of dephasing processes.

\section{Heat transport in driven $XXZ$ spin chains without dephasing} \label{nodephasing}
\begin{figure}
\begin{center}
\hspace{1.5cm}
\includegraphics{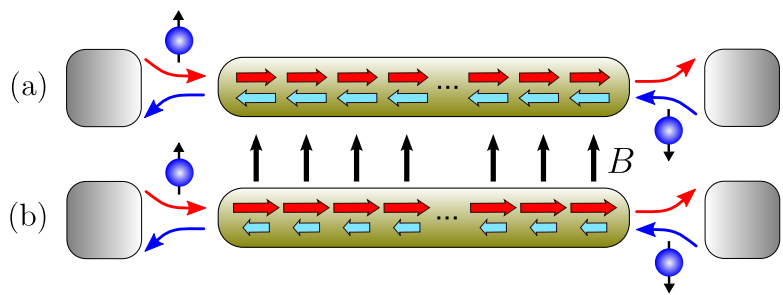}
\caption{\label{fig2} Schematic representation of heat transport in the absence of dephasing. (a) If $B=0$, the heat current of spins up injected at site $1$ (dark arrows) cancels with that of spins down injected at site $N$ (light arrows), and the net current is zero. (b) If $B>0$, the first current becomes larger than the second, and a net current emerges.}
\hspace{-1.5cm}
\end{center}
\end{figure}
\begin{figure}
\begin{center}
\hspace{1.3cm}
\includegraphics{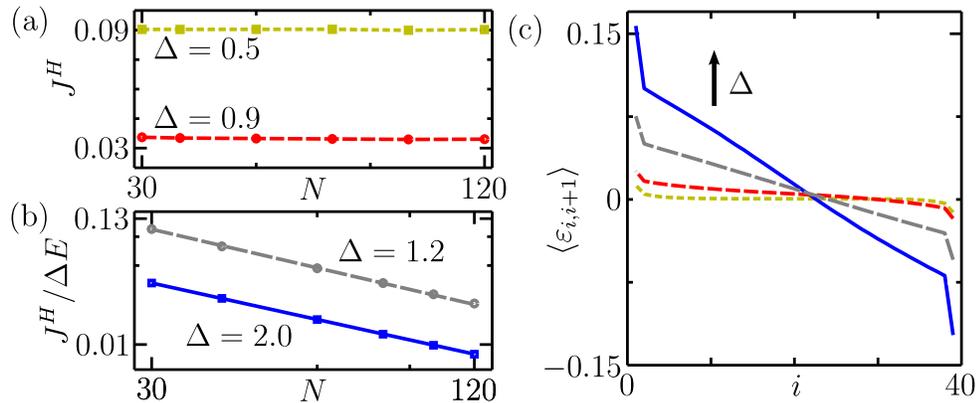}
\hspace{-1.3cm}
\caption{\label{fig3} Heat transport properties of a chain with homogeneous magnetic field and no dephasing as a function of $N$, with driving parameters $f=0.1$ and $\Gamma=1$, for different interaction strengths. (a) Heat current for $\Delta=0.9$ ($B=0.5$) and $\Delta=0.5$ ($B=1.0$); the lines are guides to the eye. (b) Log-log plot of the heat current for $\Delta>1$. The lines correspond to the fit $J^{\rm H}/\Delta E=\kappa/(N-4)^\alpha$. For $\Delta=2.0$, $\alpha=0.98$ and $\kappa=0.85$; for $\Delta=1.2$, $\alpha=1.02$ and $\kappa=2.95$. (c) Corresponding energy profiles for $N=40$, where $\Delta$ increases in the direction of the arrow.}
\end{center}
\end{figure}

Initially we consider a system without dephasing processes. We note that if the magnetic field is absent ($B=0$), there is no net heat flow through the chain, in spite of the existence of a finite spin current. This can be understood qualitatively by considering the strongly-driven case $f=1$, in which only spins up are injected at site $1$ and only spins down are injected at site $N$ (the argument is easily extended to $f<1$). At each site of the chain, the total heat current is the superposition of two contributions: one current flowing from right to left, carrying energy associated with the spins down injected at site $N$, and other flowing from left to right, carrying energy associated with the spins up injected at site $1$. Without any energetic asymmetry in the system, the contributions cancel with each other since there is no preference for any $z$ direction, as illustrated in Fig. \ref{fig2}(a). If a homogenous magnetic field is turned on, an asymmetry is established, and a total heat current emerges. If $B>0$ (as in the present work), the left-to-right flow carries energy of spins aligned with the field, so it becomes larger than the right-to-left flow. The net heat current is thus positive, as schematized in Fig. \ref{fig2}(b).\\The vanishing of the heat current at $\gamma=0$ and $B=0$ can be formally demonstrated by considering the symmetries of the Lindblad master equation~\eqref{master_eq} with $XXZ$ Hamiltonian and the boundary driving of Eq.~\eqref{driv_oper}, as discussed in Ref. \cite{popkov2013njp}. In this case, the NESS of the system is invariant under the transformation $U=\Omega^{\alpha}R$, where $\Omega^{\alpha}=\sigma_1^{\alpha}\otimes\sigma_2^{\alpha}\otimes\cdots\otimes\sigma_N^{\alpha}$, $\alpha=x,y$, and $R$ is a reflection operator, so $R(A_1\otimes B_2\otimes\cdots\otimes C_N)=(C_1\otimes\cdots\otimes B_{N-1}\otimes A_N)R$ \cite{popkov2013njp}. This means that
\begin{equation} \label{symmetry}
\rho=U\rho U^{\dag}=\Omega^{\alpha}R\rho R\Omega^{\alpha}.
\end{equation}
Since $J_i^{\rm XXZ}$ changes sign when transformed by $U$ ($U^{\dag}J_i^{\rm XXZ}U=-J_i^{\rm XXZ}$), the symmetry of Eq.~\eqref{symmetry} leads to the following property \cite{popkov2013njp}
\begin{align}
\begin{split}
\fl\langle J_i^{\rm XXZ}\rangle&=\Tr(J_i^{\rm XXZ}\rho)=\Tr(J_i^{\rm XXZ}U\rho U^{\dag})=\Tr(U^{\dag}J_i^{\rm XXZ}U\rho)\\&=-\Tr(J_i^{\rm XXZ}\rho)=-\langle J_i^{\rm XXZ}\rangle\quad\rightarrow\quad\langle J_i^{\rm XXZ}\rangle=0.
\end{split}
\end{align}
So in the absence of dephasing and magnetic field, both components of the heat current vanish, and $\langle J_i^{\rm H}\rangle=0$ for all $i$.\\We now discuss the results of heat transport for $B>0$. The first important observation is that the spin current remains unmodified if a homogeneous magnetic field is present in the system. This occurs because the effect of the magnetic field is to vary the relative energies of the sectors of fixed spin quantum number in the spectrum of the model\footnote{Each quantum number is an eigenvalue of the total spin in $z$ direction $\sum_i\sigma_i^z$, which is conserved in the $XXZ$ model.}, while leaving the internal structure of each sector unchanged. This does not affect the spin transport through the chain, which is governed by processes that conserve the number of spin excitations. The second important point is that similarly to the spin current, $\langle J_i^{\rm XXZ}\rangle$ is independent of the magnetic field. This can be seen by rewriting Eq. \eqref{xxz} in terms of $\sigma^+$ and $\sigma^-$; for example, the first term is
\begin{equation} \label{noeffect_mag_ener}
\sigma_{i-1}^y\sigma_i^z\sigma_{i+1}^x-\sigma_{i-1}^x\sigma_i^z\sigma_{i+1}^y=2i(\sigma_{i-1}^-\sigma_i^z\sigma_{i+1}^+-\sigma_{i-1}^+\sigma_i^z\sigma_{i+1}^-).
\end{equation}
The resulting terms are spin-conserving, so they only involve processes within each spin sector separately. A magnetic field does not affect expectation values of spin-conserving operators like $J_i^{\rm XXZ}$. Therefore for any homogeneous magnetic field in the absence of dephasing, $\langle J_i^{\rm XXZ}\rangle=0$, and the heat current becomes homogeneous through the chain and possesses all the features of the spin current, since $\langle J_i^{\rm H}\rangle\equiv J^{\rm H}=J^{\rm B}=BJ^{\rm S}$. Note that this is reflected in the absence of spatial imbalance of the energy density inducing $J_i^{\rm XXZ}$, i.e., $\langle h_{i,i+1}\rangle$ is homogeneous through the system.\\It is important to emphasize that the symmetry of Eq.~\eqref{symmetry} specifically applies to the Lindblad driving scheme of Eq.~\eqref{driv_oper}. Different simple driving configurations might lead to similar symmetries, resulting in vanishing particle or energy currents \cite{popkov2013njp}. Nevertheless, for other types of driving, these symmetries do not apply, leading to contributions of both $J^{\rm B}$ and $\langle J_i^{\rm XXZ}\rangle$ to the total current even in the dephasing-free case, and thus to different transport regimes than those observed in this work. \\We now study the transport properties of the dephasing-free chain at weak driving and different interaction strengths $\Delta$. Observing the scaling of the current with the size of the system, we can identify different transport regimes. As shown in Fig. \ref{fig3}(a) for $\Delta<1$, the heat current remains constant as $N$ increases, indicating ballistic transport. This picture is reinforced by the corresponding nearly flat energy profiles presented in Fig. \ref{fig3}(c). A different situation is observed for $\Delta>1$; as seen in Fig. \ref{fig3}(b), the heat currents decrease with the size of the system, and as shown in Fig. \ref{fig3}(c), the corresponding energy profiles have a constant finite gradient away from the boundaries of the chain. This behavior is characteristic of diffusive transport, in which the current is given by a diffusion equation 
\begin{equation} \label{fourier}
J^{\rm H}=\kappa_{\rm H}\nabla E\qquad\text{with}\qquad\nabla E=\frac{\Delta E}{N-4}.
\end{equation}
This is the discrete version of Fourier's law, with $\kappa_{\rm H}$ the heat conductivity, $\nabla E=\langle \varepsilon_{i,i+1}\rangle-\langle\varepsilon_{i-1,i}\rangle=\langle b_{i,i+1}\rangle-\langle b_{i-1,i}\rangle$ the constant energy gradient in the bulk and $\Delta E=\langle \varepsilon_{N-2,N-1}\rangle-\langle\varepsilon_{2,3}\rangle=\langle b_{N-2,N-1}\rangle-\langle b_{2,3}\rangle$ the energy difference across the chain (excluding the boundaries). This equation indicates that the heat current decays with the size of the system as $J^{\rm H}/\Delta E\sim N^{-1}$, which is found to agree very well with the current scaling shown in Fig. \ref{fig3}(b). Thus for the interactions $\Delta>1$ considered, the diffusion equation \eqref{fourier} is satisfied. In addition, we observe that
\begin{equation}\label{difu_no_deph}
J^{\rm H}=BJ^{\rm S}=\kappa_{\rm H}(\langle b_{i,i+1}\rangle-\langle b_{i-1,i}\rangle)=\frac{B}{2}\kappa_{\rm H}(\langle\sigma_{i+1}^z\rangle-\langle\sigma_{i-1}^z\rangle)=B\kappa_{\rm H}\nabla S.
\end{equation}
This means that the heat and spin conductivities are equal, i.e. $\kappa_{\rm H}=\kappa_{\rm S}$.\\The results shown in Fig. \ref{fig3} indicate that the critical point separating ballistic and diffusive transport regimes is very close to the isotropic point. In fact, since the heat current of the present configuration is entirely given by the magnetic field and the spin current, the critical point corresponds to $\Delta=1$, which separates the ballistic and diffusive regimes of spin transport \cite{gobert2005real,langer2009real,benenti2009charge}. This result contrasts with previous analysis of heat transport through chains driven by external reservoirs, which indicate ballistic transport for the non-interacting case only \cite{wu2011prb} and for interactions $\Delta<1.6$ \cite{michel2008prb}. It also differs from the commonly-accepted result of finite thermal Drude weight and ballistic heat transport for all interactions strengths $\Delta$ \cite{langer2011prb,zotos1997prb}, supported by integrability arguments. Nevertheless, the introduction of external reservoirs can break the integrability of the system in the most general case\footnote{Up to now, only the non-interacting case $\Delta=0$ of the boundary-driven configuration considered here has been shown to be integrable \cite{prosen2008third,prosen2010exact}; it is not known whether interacting cases are integrable too.} \cite{wu2011prb,popkov2012pre}, so it is not expected that the picture of ballistic transport for all $\Delta$ maintains its validity in the non-equilibrium configuration considered in the present work.\\It is also important to note that the ground state phase diagram of the system does not determine the nature of the (spin or heat) transport response seen here, since the equilibrium critical points strongly depend on the magnetic field \cite{takahashi}, while the interaction separating the different transport regimes is field-independent ($\Delta=1$). Thus, the transport properties of the chain are determined by the structure of its eigenspectrum, which qualitatively changes from being continuous for $\Delta<1$ to consisting of separated bands for $\Delta>1$ \cite{Sutherland}.\\
\begin{figure}
\begin{center}
\hspace{1cm}
\includegraphics{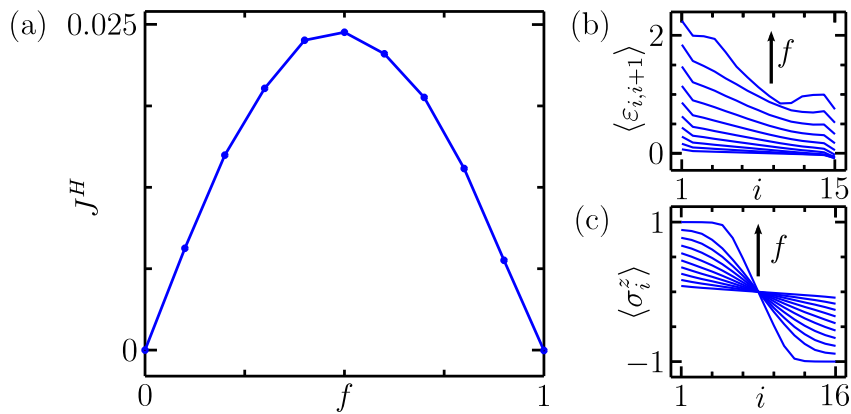} 
\hspace{-1cm}
\caption{\label{fig4} Transport properties of the strongly-interacting case $\Delta=1.5$ and driving parameters $f=0.1,0.2,\ldots,1.0$, in the absence of dephasing; $B=0.5$, $\Gamma=1$ and $N=16$. (a). Heat current. (b) Profiles of energy. (c) Magnetization. In (b) and (c), the increase of $f$ is indicated by the arrows.}
\end{center}
\end{figure}
Now we focus on the strongly-interacting regime, and obtain the heat current of the system for several drivings. Since the convergence to the NESS is slow at large driving $f\approx1$ \cite{we,benenti2009charge}, we consider systems of small size, namely $N=16$. Nevertheless, test calculations for larger systems present a similar behavior. The results are shown in Fig. \ref{fig4}. For weak driving, the heat current increases linearly with $f$ (see Fig. \ref{fig4}(a)), decreases with the size of the system (as shown in Fig. \ref{fig3}(a) for $f=0.1$), and the energy profiles have a homogeneous gradient in the bulk (see Fig. \ref{fig4}(b)), indicating diffusive transport. As the driving increases beyond $f\approx0.5$, the current starts to decrease. This large imbalance between injection and ejection of spin excitations at the boundaries favors the occupation of flat bands of bound states with low mobility (which only exist for $\Delta>1$), making it harder for the excitations to move through the chain and thus diminishing the current; this mechanism is described in Ref. \cite{we}. At the strongest driving $f=1$ the heat current vanishes, corresponding to insulating behavior (in fact it is exponentially suppressed with $N$, a tendency proved for the spin current \cite{prosenprl2011}). This phenomenon of decreasing transport with increasing driving, non-existent for weak interactions, is known as negative differential conductivity (NDC), and has been previously observed in the heat transport of $XY$ chains with $B>0$ \cite{prosen2010exact} and in the spin current through an $XXZ$ chain with $B=0$ \cite{we,benenti2009charge}. In the latter case, the insulating behavior at very strong driving results from the appearance of ferromagnetic domains of almost entirely-polarized spins at the edges of the chain, which strongly inhibit spin flips. Since the presence of a homogeneous magnetic field does not modify the spin transport features, the existence of ferromagnetic domains is maintained, as shown in Fig. \ref{fig4}(c), leading to the absence of both spin and heat currents at $f=1$.

\section{Effects of dephasing on heat transport for weak interactions} \label{depheffects1}
\begin{figure}
\begin{center}
\hspace{1cm}
\includegraphics{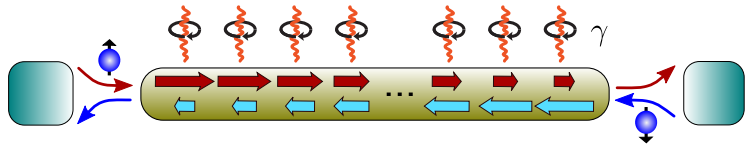}
\hspace{-1cm}
\caption{\label{fig5} Effect of dephasing on the counter-propagating heat currents coming from each boundary. When reaching each site, the currents have been attenuated by a different amount of dephasing processes, so their local magnitudes are distinct.}
\end{center}
\end{figure}

\begin{figure}
\begin{center}
\hspace{1cm}
\includegraphics{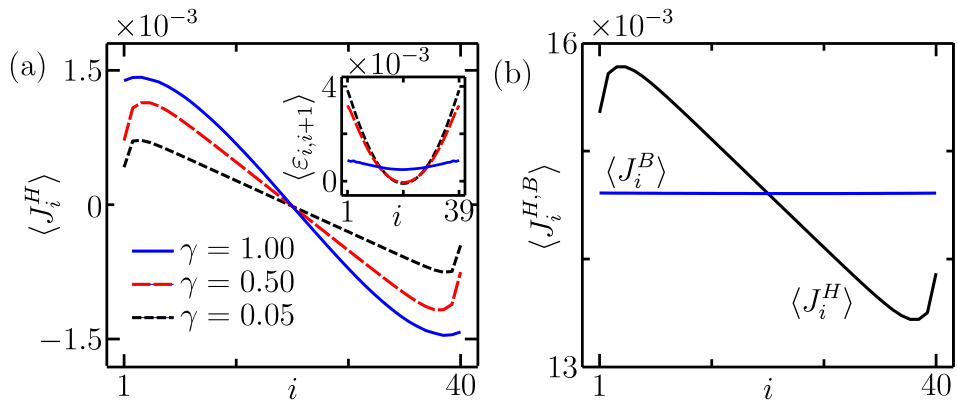} 
\hspace{-1cm}
\caption{\label{fig6} Effects of dephasing on the heat transport of the weakly-interacting regime. (a) Heat current profile for $B=0$ and different dephasing rates. The heat conductivities for $\gamma=1.00,0.5\text{ and }0.05$ are $\kappa_{\rm XXZ}=1.997(12),3.983(8)\text{ and }38(3)$, respectively. The uncertainties correspond to the standard deviation of the values of $\langle J_i^{\rm XXZ}\rangle/\nabla E_i^{\rm XXZ}$ obtained in the bulk of the chain. Inset: Corresponding energy profiles. (b) Heat current profile for $B=1$ and $\gamma=0.5$. Both panels correspond to $N=40$, $\Delta=0.5$, $f=0.1$ and $\Gamma=1$.}
\end{center}
\end{figure}

\begin{figure}
\begin{center}
\includegraphics{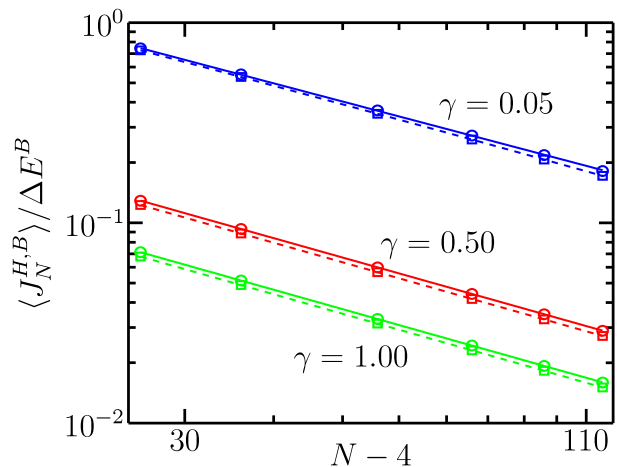}
\caption{\label{fig7} Total output heat current $\langle J_N^{\rm H}\rangle$ (\scalebox{0.6}{$\square$}) and its contribution from the magnetic field $J^{\rm B}$ ($\circ$) as a function of $N$, for $B=1$, $f=0.1$, $\Gamma=1$, $\Delta=0.5$ and dephasing rates $\gamma=0.05,0.50,1.00$. The dashed lines are guides to the eye for $J_{1}^{\rm H}$, while the solid lines are fits of the results of $J^{\rm B}$ to the equation $J^{\rm B}/\Delta E^{\rm B}=\kappa_{\rm B}(N-4)^{-\alpha}$, with $\Delta E^{\rm B}=\langle b_{N-2,N-1}\rangle-\langle b_{2,3}\rangle$. For $\gamma=0.05,0.50,1.00$, $\alpha=0.94,1.00,1.00$ and $\kappa_{\rm B}=15.85,3.35,1.86$, respectively. Note that the conductivity values differ from those of $\kappa_{\rm XXZ}$, which are field-independent (see Fig. \ref{fig6}).}
\end{center}
\end{figure}

In this section we discuss how dephasing processes affect the heat transport properties in the weakly-interacting regime. It is illustrative to study first the system in the absence of magnetic field ($B=0$), where the local heat current is entirely given by $\langle J_i^{\rm XXZ}\rangle$ (since $J^{\rm B}=0$). We consider again the picture of two currents flowing in opposite directions, associated with spin excitations injected at each boundary. In this case, the currents do not cancel with each other. Instead, they reach each site of the system with unequal magnitudes, depending on the attenuation they have experienced due to dephasing processes along the different distances they have covered, as depicted in Fig. \ref{fig5}. As a result, the net current in the left half on the chain is positive, and in the right half is negative. This means that heat flows from the boundaries towards the center of the chain. In Fig. \ref{fig6}(a) we show the heat currents for $\Delta=0.5$ and different dephasing rates, where the described behavior is observed. The corresponding energy profiles are shown in the inset of Fig. \ref{fig6}(a), indicating that the center of the chain is the region of lowest energy. In addition, we note that the amplitude of the local currents monotonically decreases with $\gamma$, and that as $\gamma$ increases, so does the amount of lost energy, as evidenced from the energy profiles.\\
Importantly, the heat current of this configuration satisfies a local diffusion equation
\begin{equation}
\langle J_i^{\rm H}\rangle=\langle J_i^{\rm XXZ}\rangle=\kappa_{\rm XXZ}\nabla E_i^{\rm XXZ},\quad\nabla E_i^{\rm XXZ}=\langle\varepsilon_{i,i+1}\rangle-\langle\varepsilon_{i-1,i}\rangle=\langle h_{i,i+1}\rangle-\langle h_{i-1,i}\rangle,
\end{equation}
where $E_i^{\rm XXZ}$ denotes the local energy density corresponding to $XXZ$ interactions. This is verified from our simulations, which indicate that the ratio $\langle J_i^{\rm XXZ}\rangle/\nabla E_i^{\rm XXZ}$ is indeed constant in the bulk of the system. Also the obtained conductivity is independent of the size of the chain, as expected for a diffusive conductor. For example, for the parameters of Fig. \ref{fig6}(a) and $\gamma=1$, the conductivities obtained for $N=40,80\text{ and }100$ are $\kappa_{\rm XXZ}=1.997(12),1.996(7)\text{ and }1.996(7)$, respectively. Therefore the net effect of dephasing is to induce diffusive heat transport from the boundaries to the center of the system.\\Next we study the heat transport through the spin chain when $B>0$ and $\gamma>0$, where both $\langle J_i^{\rm XXZ}\rangle$ and $\langle J_i^{\rm B}\rangle$ are non-zero. Recall the characteristic features of each: $\langle J_i^{\rm B}\rangle=J^{\rm B}$ is homogeneous through the chain, it has the same properties as the spin current and is proportional to the magnetic field, while $\langle J_i^{\rm XXZ}\rangle$ is spatially dependent due to dephasing, and is field-independent. In Fig. \ref{fig6}(b) we show the profile of the total heat current $\langle J_i^{\rm H}\rangle=J^{\rm B}+\langle J_i^{\rm XXZ}\rangle$ for a particular set of system parameters. The profile is anti-symmetrically centered around $J^{\rm B}$, and acquires its space dependence from $\langle J_i^{\rm XXZ}\rangle$. Near the boundaries, the dephasing processes give energy to the system due to the interplay with the driving, so $\langle J_i^{\rm H}\rangle<\langle J_{i+1}^{\rm H}\rangle$. Away from the boundaries, dephasing induces energy losses from the system, so $\langle J_i^{\rm H}\rangle>\langle J_{i+1}^{\rm H}\rangle$.\\A fundamental point to note is that as stated in earlier studies \cite{znidaric2010dephasing,znidaric2011solvable}, dephasing induces a non-equilibrium quantum phase transition from ballistic ($\gamma=0$) to diffusive ($\gamma>0$) spin transport in the weakly-interacting regime. Therefore both the spin current $J^{\rm S}$ and the heat current $J^{\rm B}$ satisfy a diffusion equation for finite dephasing similar to Eq. \eqref{difu_no_deph}, which reads
\begin{equation} \label{difu_jb}
J^{\rm B}=BJ^{\rm S}=\kappa_{\rm B}\nabla E_i^{\rm B},\qquad\nabla E_i^{\rm B}=\langle b_{i,i+1}\rangle-\langle b_{i-1,i}\rangle=B\nabla S,
\end{equation}
with $E_i^{\rm B}$ denoting the local energy density associated with the magnetic field. Both $J^{\rm S}$ and $J^{\rm B}$ are determined by the magnetization imbalance through the system, and have the same conductivity, $\kappa_{\rm B}=\kappa_{\rm S}$. This leads to a scaling of $J^{\rm B}$ with the size of the system like that indicated in Eq. \eqref{fourier}. As shown in Fig. \ref{fig7} this agrees very well with our results.\\The properties of $J^{\rm B}$ and $\langle J_i^{\rm XXZ}\rangle$ indicate that for finite dephasing rates and magnetic fields, the total heat current at any site $i$ of the bulk is the sum of two diffusive contributions. Each one is determined by the gradient of a different component of the local energy density, and has a different conductivity (note the values of $\kappa_{\rm B}$ and $\kappa_{\rm XXZ}$ of Figs. \ref{fig6} and \ref{fig7}). So we can write
\begin{equation} \label{total_difu}
\langle J_i^{\rm H}\rangle=J^{\rm B}+\langle J_i^{\rm XXZ}\rangle=\kappa_{\rm B}\nabla E_i^{\rm B}+\kappa_{\rm XXZ}\nabla E_i^{\rm XXZ}.
\end{equation}
Our results indicate that the total heat current does not fulfill a diffusion equation, i.e., that an expression such as $\langle J_i^{\rm H}\rangle=\kappa_{\rm H}\nabla(E_i^{\rm B}+E_i^{\rm XXZ})=\kappa_{\rm H}\nabla E_i^{\rm H}$, with $E_i^{\rm H}$ representing the total local energy density, is not valid \footnote{This is seen by noting that the ratio $\langle J_i^{\rm H}\rangle/\nabla E_i^{\rm H}$ is not homogeneous through the bulk of the system, which results from having $\kappa_{\rm B}\neq\kappa_{\rm XXZ}$.}. Nevertheless, since it consists of the addition of two diffusive components, it is clear that the existence of dephasing processes changes the nature of the heat transport through the system for $\Delta<1$. Therefore, similarly to spin transport \cite{znidaric2010dephasing,znidaric2011solvable}, heat transport shows a dephasing-induced non-equilibrium phase transition between ballistic and diffusive behaviors in the weakly-interacting regime. A similar result has been reported recently in the non-interacting case $\Delta=0$, for a different driving scheme and smaller systems \cite{manzano2012pre}.\\Importantly, observe that for any combination of incoherent processes, interaction strength and system size, we can always find a magnetic field weak enough to cause $|\langle J_N^{\rm XXZ}\rangle|>J^{\rm B}$, and thus that $\langle J_N^{\rm H}\rangle<0$. Since $\langle J_{1}^{\rm H}\rangle$ is always positive (both $J^{\rm B}>0$ and $\langle J_1^{\rm XXZ}\rangle>0$), this corresponds to a configuration in which the chain absorbs energy at both boundaries and dissipates it by dephasing processes. Some examples of these case are shown in Section \ref{inversion}. However, in the present work we mostly focus on cases where $\langle J_N^{\rm H}\rangle>0$, in which the system delivers some energy to the right reservoir absorbed from the left reservoir. For such cases we observe the behavior of the total output current $\langle J_N^{\rm H}\rangle$; this is shown in Fig. \ref{fig7} as a function of the size of the system for different dephasing rates. Since $J^{\rm B}$ is largely dominant over $\langle J_N^{\rm XXZ}\rangle$ for the parameters of Fig. \ref{fig7}, the corresponding output heat current $\langle J_N^{\rm H}\rangle$ scales with $N$ in a form very similar to $J^{\rm B}$.\\Finally, it is known that for $\Delta<1$ the spin current decreases monotonically with the dephasing rate \cite{we,znidaric2011solvable,znidaric2010dephasing}; the same thus happens for $J^{\rm B}$, as seen in Fig. \ref{fig7}. Since this is also the case for $\langle J_i^{\rm XXZ}\rangle$ (see Fig. \ref{fig6}), the total heat current $\langle J_i^{\rm H}\rangle$ decreases monotonically with $\gamma$ at any site $i$ of the chain. This shows that dephasing only degrades heat transport in the weakly-interacting regime.

\section{Effects of dephasing on heat transport for strong interactions} \label{depheffects2}

In this Section we analyze how dephasing affects heat transport in the strongly-interacting regime $|\Delta|>1$. We focus on two main effects: the suppression of NDC and dephasing-enhanced heat transport. 

\begin{figure}
\begin{center}
\hspace{2cm}
\includegraphics{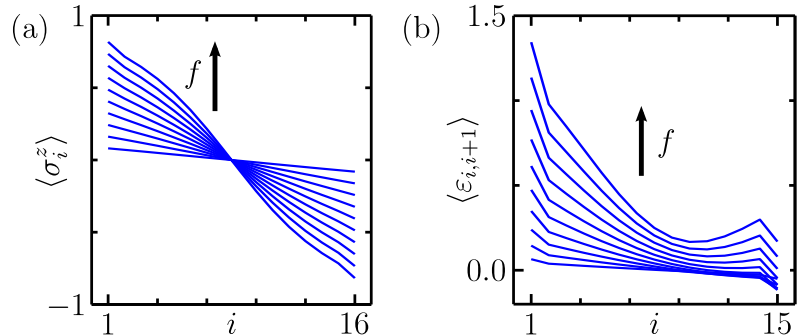} 
\caption{\label{fig8} Spin profiles (a) and energy profiles (b) for the same parameters of Fig. \ref{fig4}, but with dephasing rate $\gamma=0.3$. The arrows indicate the increase of the driving $f$, with $f=0.1,0.2,\ldots,1.0$.}
\hspace{-2cm}
\end{center}
\end{figure}

\subsection{Suppression of NDC} \label{inversion}

The NDC effect presented in Fig. \ref{fig4} is modified by dephasing. This is already made clear by considering how dephasing affects the spin and energy profiles, as shown in Fig. \ref{fig8} for several drivings and a particular rate $\gamma$. The ferromagnetic domains inhibiting spin transport at large driving do not appear once $\gamma>0$, as seen in Fig. \ref{fig8}(a), leading to diffusive spin conduction for the entire driving range \cite{we}. The energy profiles, shown in Fig. \ref{fig8}(b), feature a more complicated behavior. At driving $f\lesssim0.3$, their form is similar to that of the case without dephasing (see Fig. \ref{fig4}), suggesting heat conduction from the left to the right reservoir. At driving $f\gtrsim0.3$, they indicate heat flows from both boundaries of the chain towards the bulk. This means that when the driving increases beyond a certain value, the output heat current is reversed.\\
\begin{figure}
\begin{center}
\raggedleft
\includegraphics{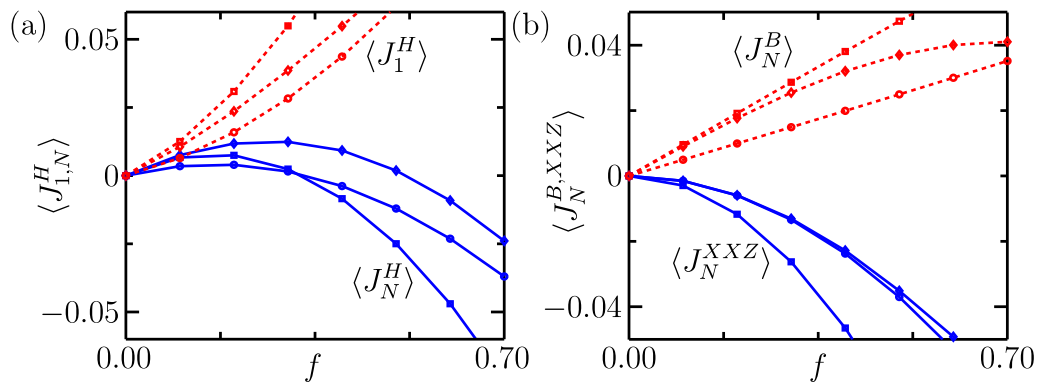}
\caption{\label{fig9} Heat current and its components as a function of the driving $f$, for dephasing rates $\gamma=0.05$ ($\diamond$), $\gamma=0.30$ (\scalebox{0.6}{$\square$}) and $\gamma=2.00$ ($\circ$). The common parameters of the simulations are $\Gamma=1$, $B=0.5$, $\Delta=1.5$ and $N=16$. (a) Total input ($\langle J_{1}^{\rm H}\rangle$) and output ($\langle J_N^{\rm H}\rangle$) heat currents. (b) Components of the output heat current $J^{\rm B}$ and $\langle J_N^{\rm XXZ}\rangle$. Note that for a fixed driving, both currents have a non-monotonic behavior with $\gamma$; for the plotted cases, their largest magnitude occurs for the intermediate dephasing rate $\gamma=0.3$.}
\end{center}
\end{figure}
Now we analyze the heat current. As in the case of weak interactions and $\gamma>0$, both contributions to the total heat current satisfy a diffusion equation with different conductivities, so Eq. \eqref{total_difu} remains valid\footnote{It is known from earlier studies that the spin current shows diffusive behavior for $\Delta>1$, at any driving with dephasing and at weak driving without dephasing \cite{we,znidaric2010dephasing}; the same thus applies to $J^{\rm B}$. The diffusivity of $\langle J_i^{\rm XXZ}\rangle$ is proved as before: by observing that the ratio $\langle J_i^{\rm XXZ}\rangle/\nabla E_i^{\rm XXZ}$ is constant in the bulk, and independent of $N$.}. We focus on the boundary currents, which are shown in Fig. \ref{fig9}(a) for different drivings and dephasing rates. While the input current $\langle J_{1}^{\rm H}\rangle$ remains positive and increases its magnitude with $f$, the output current $\langle J_N^{\rm H}\rangle$ shows a completely different behavior. Initially, for very weak driving, it is positive and increases with $f$. Then it decreases with small increments of the driving. For even larger driving, its direction is reversed, so heat starts flowing to the left from the right boundary.\\To understand how this behavior of $\langle J_N^{\rm H}\rangle$ arises, we observe its components  $J^{\rm B}$ and $\langle J_N^{\rm XXZ}\rangle$ separately, which are shown in Fig. \ref{fig9}(b). The component $J^{\rm B}$, always flowing to the right, features NDC for small dephasing rates (such as $\gamma=0.05$ for the parameters of Fig. \ref{fig9}), but now it presents a finite conductivity at $f=1$, a behavior not shown in Fig. \ref{fig9}(b). For larger rates (for example, $\gamma=0.3$ and $2.0$), NDC disappears completely from $J^{\rm B}$, its amplitude monotonically increasing with $f$. In contrast, the component $\langle J_N^{\rm XXZ}\rangle$ always flows to the left, and grows monotonically and rapidly as the driving gets stronger. This occurs due to the favored population of high-energy bound states as $f$ increases, resulting in a larger amount of energy to be dissipated and thus in a larger heat flow from the boundaries to the central part of the chain due to dephasing processes. The addition of both components leads to the observed behavior of the total current at the right edge of the chain $\langle J_N^{\rm H}\rangle$. Depending on which contribution is dominant, the system will feature heat flow from both boundaries towards the bulk, or a net flow to the right reservoir.

\subsection{Dephasing-enhanced heat transport} \label{enhancement}
\begin{figure}
\begin{center}
\hspace{1cm}
\includegraphics{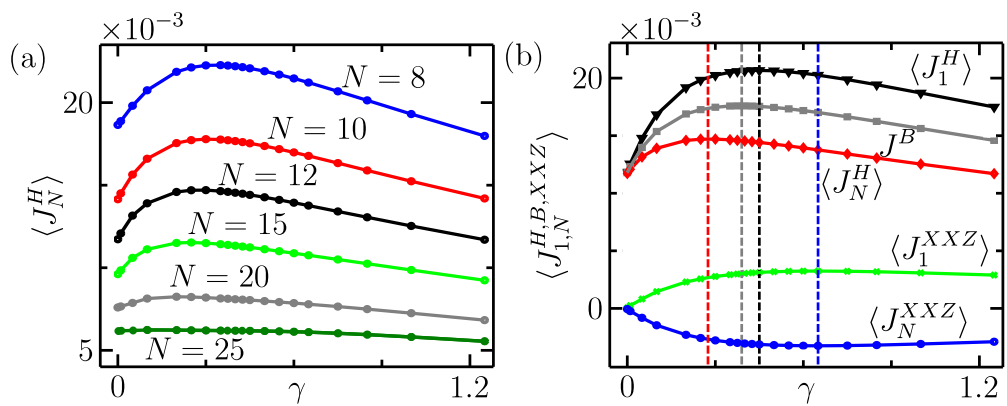}
\hspace{-1cm}
\caption{\label{fig10} Dephasing-enhanced heat transport at weak driving, with $f=0.1$, $\Gamma=1$, $B=1$ and $\Delta=2.0$. (a). Heat current at site $N$ as a function of $\gamma$, for several system sizes $N\leq25$. (b). Heat currents at the boundaries and their field-dependent and independent contributions as a function of $\gamma$, for $N=12$. Each vertical dashed line indicates the optimal dephasing rate $\gamma_{\text{opt}}$ of the current plotted in the same color; $\gamma_{\text{opt}}\approx0.28,0.39,0.45,0.65$ for $\langle J_{N}^{\rm H}\rangle,J^{\rm B},\langle J_1^{\rm H}\rangle,\langle J_{1,N}^{\rm XXZ}\rangle$, respectively.}
\end{center}
\end{figure}
In earlier work we found that for any driving, the spin current can be significantly enhanced by dephasing in the strongly-interacting regime \cite{we}. This result also applies to the field-dependent heat current component $J^{\rm B}$, as seen in Fig. \ref{fig9}(b). Now we consider whether a similar effect of environment assistance exists for the total heat current in the conventional scenario of unidirectional transport. For different values of $N$, the total output heat current $\langle J_N^{\rm H}\rangle$ is shown as a function of $\gamma$ in Fig. \ref{fig10}(a). We find that for chains of small size and moderate dephasing rates, the output current can be considerably larger than that of the case $\gamma=0$ (for example, when $N=8$ the enhancement of the current is $\approx$ 19\%), indicating environment-assisted total heat transport. For large dephasing rates, the effect is no longer visible due to the degradation of coherent dynamics by the frequent perturbation of the environment (e.g. Zeno effect \cite{breuer}) and the bulk energy dissipation.\\
\begin{figure}[t]
\begin{center}
\hspace{1cm}
\includegraphics{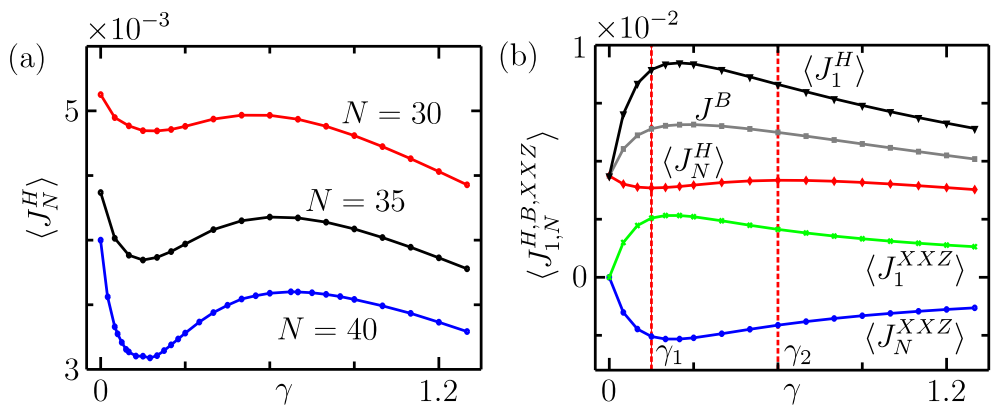}
\hspace{-1cm}
\caption{\label{fig11} Multiple change of monotonicity of the heat transport as a function of the dephasing rate $\gamma$, with $f=0.1$, $\Gamma=1$, $B=1$ and $\Delta=2.0$. (a). Heat current at site $N$ as a function of $\gamma$, for different system sizes $N\geq30$. (b). Heat currents at the boundaries and their field-dependent and independent contributions as a function of $\gamma$, for $N=35$. The vertical dashed lines correspond to the dephasing rates $\gamma_1\approx0.15$ and $\gamma_2\approx0.6$, which indicate the changes on monotonicity for $N=35$.}
\end{center}
\end{figure}
The components of the heat current at both boundaries for $N=12$ are shown in Fig. \ref{fig10}(b). We see that in contrast to the weakly-interacting regime (see Fig. \ref{fig6}(a)), the boundary currents $\langle J_{1,N}^{\rm XXZ}\rangle$ (and in general the currents $\langle J_i^{\rm XXZ}\rangle$ for all $i$) feature initial increase and subsequent decay with $\gamma$ for strong interactions. So the two components of the total heat current show enhancement for moderate dephasing rates. Since both flow in opposite directions in the output, they compete with each other to determine the total heat delivered to the right reservoir. The corresponding optimal dephasing $\gamma_{\text{opt}}$ and the rates of increase and degradation with $\gamma$ are different for both; this leads to an intricate interplay between $J^{\rm B}$ and $\langle J_N^{\rm XXZ}\rangle$ to determine the total output heat current. As clearly seen in Fig. \ref{fig10}(b), dephasing-assisted heat transport compared to the dephasing-free case exists as long as $J^{\rm B}$ is the dominant contribution to the total current over $\langle J_N^{\rm XXZ}\rangle$, and as $J^{\rm B}$ features environment assistance itself.\\Finally, note that in contrast to spin transport, where dephasing assistance compared to the case $\gamma=0$ occurs for any system size \cite{we}, the heat transport only displays such an enhancement for small chains. For example, for the parameters used in Fig. \ref{fig10}, a chain with $N=25$ already experiences enough energy dissipation in the bulk to induce a barely significant current enhancement (1.3\%). Nevertheless, a different assistance effect emerges for larger systems, as shown in Fig. \ref{fig11}(a). For small dephasing rates, the output heat current $\langle J_N^{\rm H}\rangle$ is initially degraded compared to the dephasing-free case. But from a rate $\gamma_1$, increasing $\gamma$ leads to a significant current enhancement, until degradation is induced again after a rate $\gamma_2$ due to the Zeno effect. From Fig. \ref{fig11}(b) we can see why this multiple change of monotonicity, absent in the input current $\langle J_1^{\rm H}\rangle$, occurs on large chains. At small dephasing rates $\langle J_N^{\rm XXZ}\rangle$ is enhanced (negatively) faster than $J^{\rm B}$, so the total current is initially degraded. At $\gamma_1$ the situation is inverted, leading to an increase of $\langle J_N^{\rm H}\rangle$ with $\gamma$. This tendency continues for moderate dephasing rates up to $\gamma_2$, even when both $J^{\rm B}$ and $\langle J_N^{\rm XXZ}\rangle$ are degraded by $\gamma$, since the latter decreases faster than the former. This shows that a beneficial role of dephasing for heat transport is also present in large systems.\\ 
\begin{figure}[t]
\begin{center}
\hspace{1cm}
\includegraphics{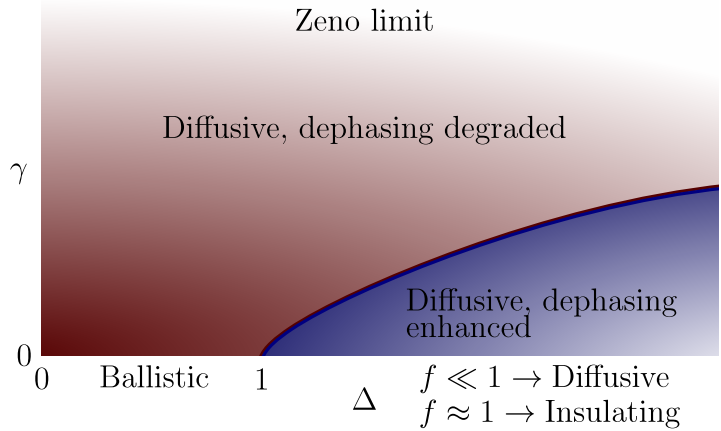}
\hspace{-1cm}
\caption{\label{fig12} Schematic diagram of the spin and heat transport regimes; higher color intensities correspond to larger currents. In the absence of dephasing ($\gamma=0$), the transport is ballistic for the weakly-interacting regime $\Delta<1$, diffusive for the strongly-interacting regime $\Delta>1$ and weak driving, and insulating for $\Delta>1$ and strong driving. When dephasing is considered ($\gamma>0$), the transport for weak interactions becomes diffusive, with the output currents decaying monotonically with $\gamma$; this corresponds to dephasing degraded transport. In contrast, for strong interactions, the transport is enhanced by moderate dephasing rates, until an optimal rate which increases with $\Delta$. As discussed in Ref. \cite{we} for spin transport, the latter behavior is expected since as the interactions get stronger, a larger environmental coupling is required to break the spin bound states; this results in the boundary shown in the figure. A similar qualitative tendency is expected for the heat transport, since the heat current and its components are enhanced by moderate dephasing. Further increase of $\gamma$ leads to current degradation, until reaching the Zeno limit, in which the $z$ degrees of freedom are frozen and the dynamics of the system is weakly dependent on $\Delta$.}
\end{center}
\end{figure}

\section{Conclusions} \label{conclu}
In the present work we have studied the heat transport in non-equilibrium quantum spin chains with a homogeneous magnetic field, and observed how it is affected by energy-dissipating local dephasing processes. The heat current results from a spin imbalance at the boundaries of the chain, induced by a Lindblad-type driving, so it corresponds to the magnetothermal response of the system. Our results are summarized in Fig. \ref{fig12}, which qualitatively describes both spin and heat transport. To obtain the steady state of the system efficiently, we simulated its time evolution under the corresponding Lindblad master equation using the matrix product based TEBD method. We observed that the heat current consists of two components: a homogeneous current proportional to the magnetic field corresponding to the heat flow carried by the spin current, and a field-independent spatially varying current arising from the $XXZ$ kinetics and the energy dissipation caused by dephasing, flowing from the boundaries towards the center of the chain.\\Initially we found that in the absence of dephasing the heat current is homogeneous through the system, and presents the same properties as the spin current. This corresponds to ballistic transport in the weakly-interacting regime $\Delta<1$, while in the strongly-interacting regime $\Delta>1$ the system shows diffusive transport for weak driving $f$, NDC as the driving increases, and insulating behavior at large driving $f\sim1$. These results hold for any finite amplitude of the magnetic field, thus being independent of the (gapless or gapped) nature of the ground state of the Hamiltonian, and instead more generally dependent on the energy eigenstructure.\\We proceeded to show that in the presence of dephasing processes, the two components of the heat current are finite and compete with each other. Depending on which one dominates, the total current at the boundaries can be positive (heat delivered from the left to the right reservoir) or of opposite signs (heat flowing towards the bulk). We showed that for finite dephasing rates, both current components satisfy a diffusion equation with different conductivities and are determined by the gradients of different components of the total energy. In the weakly-interacting regime, this corresponds to a change of the nature of heat transport, i.e., a dephasing-induced non-equilibrium phase transition between ballistic and diffusive regimes. For strong interactions, dephasing processes degrade the NDC effect, turning the system into a diffusive conductor for all drivings $f$.\\We also found that for weak interactions, the heat current decreases monotonically as the dephasing rate increases, so this coupling to the environment only degrades transport. Instead, environment-assisted spin and heat transport emerge at moderate dephasing rates in the strongly-interacting regime. While the enhancement of the spin current by dephasing exists for any size of the chain \cite{we}, that of the heat current only occurs for small systems, since dephasing dissipates more energy as the size of the system increases. Nevertheless, a beneficial role of dephasing on the total output heat current can also be observed for larger systems, when compared to configurations with a small non-zero dephasing rate.\\The mechanism underlying the transport enhancement relies on dephasing-induced transitions from flat bands of bound states to mobile bands of scattering states \cite{we}. Since this is quite generic, we expect that the results reported in this work can be experimentally observed for both spin and heat transport, and will be relevant for some cuprate materials and controllable interacting systems such as ultracold atomic gases.\\Finally we note that driving schemes alternative to that considered in the present work, such as those directly imposing an energy imbalance between the boundaries of the chain~\cite{prosen2009matrix,ajisaka2012prb}, might result in different transport properties. However, we expect dephasing-enhanced transport to emerge also in these cases at strong interactions, given the generality of its origin. These effects are currently under study and will be reported in a forthcoming publication \cite{we2}.

\ack
We acknowledge the Oxford Supercomputing Center for providing resources to perform the calculations presented in this work, and C. Greenough's group at STFC RAL for sofware support. We also acknowledge support through HECToR under the University dCSE scheme provided by NAG, carried out by Chris Goodyer. J.J. M.-A. acknowledges Departamento Administrativo de Ciencia, Tecnolog\'{i}a e Innovaci\'{o}n Colciencias for economic support, as well as Thomas Grujic and Mark Mitchison for helpful discussions. SA acknowledges support from the CCPQ flagship project (EP/J010529/1). SRC and DJ thank the National Research Foundation and the Ministry of Education of Singapore for support.\\

\bibliographystyle{unsrt}

\bibliography{mybib}

\end{document}